\documentclass[12pt,preprint]{aastex}

\usepackage{amsmath,amssymb} %Needed for bold greek math, etc.

\shorttitle{Circulation on hot Jupiters}
\shortauthors{Goodman}

\newcommand{\ud}{{\rm d}}
\newcommand{\D}{D}
\newcommand{\vk}{\boldsymbol{\hat k}}
\newcommand{\vOmega}{\boldsymbol{\Omega}}
\newcommand{\vv}{\boldsymbol{v}}
\newcommand{\vvh}{\vv_{\rm h}}

\newcommand{\bdot}{\boldsymbol{\cdot}}
\newcommand{\cross}{\boldsymbol{\times}}
\newcommand{\grad}{\boldsymbol{\nabla}}
\newcommand{\gradh}{\grad_{\rm h}}

\newcommand{\cs}{c_{\rm s}} % Sound speed
\newcommand{\BV}{Brunt-V\"ais\"al\"a\ }
\begin{document}
\title{Thermodynamics of atmospheric circulation on hot Jupiters}


\author{J. Goodman}
\affil{Princeton University Observatory, Princeton, NJ 08544}
\email{jeremy@astro.princeton.edu}

\begin{abstract}
  Atmospheric circulation on tidally-locked exoplanets is driven by
  the absorption and reradiation of heat from the host star.  They are
  natural heat engines, converting heat into mechanical energy.  A
  steady state is possible only if there is a mechanism to dissipate
  mechanical energy, or if the redistribution of heat is so effective
  that the Carnot efficiency is driven to zero.  Simulations based on
  primitive, equivalent-barotropic, or shallow-water equations without
  explicit provision for dissipation of kinetic energy and for
  recovery of that energy as heat, violate energy conservation.  More
  seriously perhaps, neglect of physical sources of drag may
  overestimate wind speeds and rates of advection of heat from the day
  to the night side.
  \end{abstract}




\goodbreak
\section{Introduction}
\cite{Showman_Menou_Cho07} have called attention to the divergence in
results among attempts to simulate atmospheric ciculation on strongly
irradiated jovian exoplanets (``hot Jupiters'').  Simulations that
allow explicitly for absorption and rejection of heat by the
atmosphere, whether by simple Newtonian-cooling schemes or by
radiative transfer, generally find prograde equatorial wind speeds
$1\mbox{-}5\,{\rm km\,s^{-1}}$
\citep{Showman_Guillot02,Burkert_etal05,Cooper_Showman05,
  Langton_Laughlin08,Showman_Cooper_etal08,Dobbs-Dixon_Lin08}.  For
comparison, the sound speed of nondegenerate molecular hydrogen is
approximately $2.4(T/10^3\,{\rm K})^{1/2}\,{\rm km\,s^{-1}}$, so many
of these winds are supersonic.  Simulations that represent uneven
heating indirectly, via prescribed lateral variations in the thickness
or depth of the circulating layer, typically have peak winds $<1\,{\rm
  km\,s^{-1}}$ \citep{Cho_etal03,Langton_Laughlin07,Cho_etal08}, and
they sometimes predict that the equatorial flow is retrograde.  In the
simulations of \cite{Cho_etal03} and \cite{Cho_etal08}, the
root-mean-square wind speed is put in by hand in the initial
conditions.  Unlike wind speeds, light curves of these planets have
been directly measured, or at least constrained, by infrared
observations
\citep{Harrington_etal06,Harrington_etal07,Cowan_etal07,Knutson_Charbonneau_etal07}.
The models generally do not predict these light curves very well: in
particular, they sometimes produce a $60\mbox{-}90^\circ$ prograde
shift of the photospheric temperature extrema with respect to the
substellar and antistellar points
\citep{Cooper_Showman05,Langton_Laughlin07}, and they sometimes
predict shifts of the wrong sign, as for the cold spot on HD187933b
\citep{Showman_Cooper_etal08}.

There may be many reasons for the disagreements among the models and
between the models and the observations.  For example, with the
exception of those by \citet{Burkert_etal05} and by
\citet{Dobbs-Dixon_Lin08}, the calculations use approximations based
in part on the presumption of subsonic flow, even though the results
sometimes violate that presumption.  The purpose of the present paper,
however, is to emphasize one aspect of the physics that has not been
adequately respected by any of the simulations: namely, that
atmospheric circulation is a natural heat engine, and that dissipation
of mechanical energy must therefore be explicitly addressed.  While it
may not be easy to decide how dissipation occurs on real planets, it
is possible that if a common model for dissipation were adopted, then
the calculations might agree better on gross features of the flow such
as peak wind speeds and light curves.  Without such a common model,
convergence to a common result seems unlikely.

\section{Production of mechanical work}
\label{sec:work}

Three of the most recent models of radiatively driven circulation on
exoplanets have been based on direct integration of the so-called
``primitive equations'' adapted from terrestrial meteorology
\citep{Showman_Cooper_etal08,
  Menou_Rauscher08,Showman_Fortney_etal08}.  Compared to previous
efforts to model the dynamics of atmospheric circulation,
\cite{Showman_Fortney_etal08} in particular have significantly
improved the model for coupling the flow to the radiation field.  Our
discussion does not do justice to those improvements, because our
concern is with the dynamical part of the model, which has much in common
with other efforts, including some of those that employ height-averaged
versions of the primitive equations: more detail is given in \S\ref{sec:discussion}.

The primitive equations are
\begin{equation}
  \label{eq:euler}
  \frac{\D\vvh}{\D t}= -\gradh\Phi-f\vk\cross\vvh,
\end{equation}
\begin{equation}
  \label{eq:hydrostatic}
  \frac{\partial\Phi}{\partial p}=\,-\,\frac{1}{\rho}\,,
\end{equation}
\begin{equation}
  \label{eq:continuity}
  \gradh\bdot\vvh+\frac{\partial\omega}{\partial p}=0,
\end{equation}
\begin{equation}
  \label{eq:entropy}
  \frac{\D s}{\D t}=\frac{q}{T}\,.
\end{equation}
The subscript ``$h$'' denotes the two horizontal components of a
vector or vectorial operator parallel to isobaric surfaces.  The
quantity $\Phi=gz$ is the gravitational potential; $\vk$ is the unit
normal to the isobars; $f\equiv 2\vOmega\bdot\vk$ is the planetary
vorticity, so that $f\vk\times\vvh$ is the projection of the Coriolis
acceleration onto the isobars; $\rho$ is the mass density;
$\omega\equiv \D p/\D t$ is the vertical ``velocity'' in pressure
coordinates; $s$ and $q$ are the entropy and radiative heating rate
per unit mass, respectively; and
\begin{equation}
  \label{eq:ddt}
  \frac{\D}{\D t}= \frac{\partial}{\partial t}+\vvh\bdot\gradh+
\omega\frac{\partial}{\partial p}
\end{equation}
is the time derivative following the flow.  Equation
\eqref{eq:hydrostatic} is a statement of hydrostatic equilibrium
perpendicular to the isobars, so that vertical accelerations have been
neglected.  This is a good approximation on large horizontal scales,
since the vertical pressure scale height is small compared to the
radius of the planet ($H_p\sim10^{-2}R$), but it precludes explicit
representation of small-scale three-dimensional turbulence, which
might be an important dissipation mechanism.  The continuity equation
\eqref{eq:continuity} allows for seepage of material across the
isobars.  \citet{Showman_Fortney_etal08} evaluate the heating rate $q$
in the entropy equation from the vertical divergence $g\partial
F/\partial p$ of the radiative flux, while
\citet{Showman_Cooper_etal08} and \citet{Menou_Rauscher08} use simpler
Newtonian schemes in which $q\propto(T-T_{\rm eq})/\tau$, where
$T_{\rm eq}$ is some prescribed reference temperature, which varies
with position, and $\tau$ is a prescribed equilibration time.
Equation \eqref{eq:entropy} is often rewritten in terms of the
temperature and the specific heat at constant pressure, $c_p$;
\citet{Showman_Fortney_etal08} rewrite it in terms of potential
temperature $\Theta\equiv T(p/p_0)^{(\gamma-1)/\gamma}$ on the
assumption that the adiabatic index $\gamma\equiv c_p/c_v$ is
constant.  What is important here is that $q$ is interpreted to
include radiative heating only; other sources of entropy such as
viscous dissipation have been neglected in the simulations that we describe.

One can derive an expression for the rate of production of mechanical
energy based on the equations above.  The first steps are
\begin{align}\label{eq:steps}
\frac{\D}{\D t}\left(\frac{v_{\rm h}^2}{2}\right)=\vvh\bdot\frac{\D\vvh}{\D t}
&= -\vvh\bdot\gradh\Phi &\mbox{from \eqref{eq:euler}}\nonumber\\
&= -\frac{\D\Phi}{\D t} +\omega\frac{\partial\Phi}{\partial p}
+\frac{\partial\Phi}{\partial t} &\mbox{using \eqref{eq:ddt}}\nonumber\\
&= -\frac{\D\Phi}{\D t} -\frac{\omega}{\rho}
+\frac{\partial\Phi}{\partial t} &\mbox{using \eqref{eq:hydrostatic}}\,.
\end{align}
Recasting $\omega/\rho=\rho^{-1}\D p/\D t$ as $\D(p\rho^{-1})/\D
t-p\D(\rho^{-1})/\D t$ and rearranging terms results in
\begin{equation}
  \label{eq:dmech0}
  \frac{\D}{\D t}\left(\frac{1}{2}\vvh^2+\Phi+\frac{p}{\rho}\right)
-\frac{\partial\Phi}{\partial t} = p\frac{\D}{\D t}\frac{1}{\rho}\,.
\end{equation}

The next step is to integrate eq.~\eqref{eq:dmech0} over the mass of
the atmosphere down to a depth or pressure level sufficiently great
that all time derivatives below that level can be neglected, but
sufficiently shallow so that the vertical gravity
$g\equiv\partial\Phi/\partial z$ can be treated as spatially as well
as temporally constant.  To do this for the partial time derivative of
the potential, which is evaluated at constant pressure rather than
constant depth, we use eq.~\eqref{eq:hydrostatic} to write the mass
element $\ud m\equiv \rho\ud^3\boldsymbol{r}$ as $\ud m=g^{-1}\ud p\ud
A$, where $\ud A$ is an element of area on the isobars:\footnote{We
  have assumed in eq.~\eqref{eq:dphi} that $g^{-1}\ud A$ is constant
  at a given horizontal position on an isobar as it rises or subsides.
  This would be true in a plane-parallel geometry with strictly
  constant gravity, since the curvature of the isobars would then be
  second order in the small ratio of the vertical and horizontal
  lengthscales.  But in spherical geometry, changes in $g$ and $\ud A$
  can be first order in vertical motions: for example, a spherical
  isobar of varying radius $r_p(t)$, has area $A\propto r_p^2$ and
  gravity $g_p\propto r_p^{-2}$ so that $g_p^{-1}\ud A_p\propto
  r_p^4$.  We neglect this complication since we are mainly concerned
  with the possibility of steady-state circulation, where the isobars
  would be fixed and $(\partial\Phi/\partial t)_p$ would vanish
  \emph{a fortiori}.}
\begin{equation}\label{eq:dphi}
  \int\left(\frac{\partial\Phi}{\partial t}\right)_p \ud m=
  \int\left(\frac{\partial\Phi}{\partial t}\right)_p \frac{\ud p \ud A}{g}=
  \frac{\ud}{\ud t}\int \Phi\,\frac{\ud p \ud A}{g}=
  \frac{\ud}{\ud t}\int \Phi\,\ud m=\int\frac{\D\Phi}{\D t}\, \ud m.
\end{equation}
The integral of eq.~\eqref{eq:dmech0} over the mass of the atmosphere
is therefore
\begin{subequations}\label{eq:twins}
\begin{equation}
  \label{eq:dmech1}
    \frac{\ud}{\ud t}\int\left(\frac{v_{\rm h}^2}{2}
+\frac{p}{\rho}\right)\ud m=
\int p\frac{\D}{\D t}\rho^{-1}\,\ud m\,.
\end{equation}
Using the First Law of Thermodynamics $\ud\varepsilon=T\ud
s-p\ud(\rho^{-1})$, where $\varepsilon$ is the internal energy per
unit mass, we may write eq.~\eqref{eq:dmech1} in the equivalent form
\begin{equation}
  \label{eq:dmech}
  \frac{\ud}{\ud t}\int\left(\frac{v_{\rm h}^2}{2}+\varepsilon
+\frac{p}{\rho}\right)\ud m
=\int T\frac{\D s}{\D t}\,\ud m\,.
\end{equation}
\end{subequations}
The integrand on the left side is the Bernoulli constant of the flow,
in which $\Phi$ does not appear because the work done against gravity
is balanced by the vertical component of the pressure gradient due to
the assumption of vertical hydrostatic equilibrium.

\section{Frictionless heat engines}

The integrals on the left sides of eqs.~\eqref{eq:twins} are functions
of state.  Therefore in the absence of secular changes, the time
averages of the left sides vanish.  But the integrands on the right
sides are not time derivatives of functions of state, and so their
averages will not vanish automatically.  According to equation
\eqref{eq:entropy}, the entropy of each fluid element changes only by
radiative exchange of heat; generation of heat by viscous dissipation has
not been allowed for.  Insofar as the elements that lose heat
are cooler than average (those on the night side), and those that
gain it are warmer than average (those on the day side), the righthand
side of equation \eqref{eq:dmech} should be positive.  The radiatively
driven circulation is a heat engine, continuously producing
mechanical energy by tapping the flow of heat from the day to the
night side.

Real heat engines, whether artificial or natural, have loads: the
mechanical energy that they produce is ultimately returned to heat
through an irreversible process.  Engineers minimize the dissipation
that occurs within the engine itself in order to maximize the power
transmitted to an external load.  The thermodynamic efficiency of the
engine is calculated as the fraction of the heat absorbed from the hot
reservoir that is converted to work rather than rejected to the cold
reservoir; in this calculation, the heat rejected by the external load
is not counted.  In self-contained natural heat engines, by contrast,
dissipation must be entirely internal.  For example, the kinetic
energy of convection is dissipated viscously, even at very high
Reynolds numbers as in stellar convection zones: turbulent cascades or
shocks, if convection is transonic, bring kinetic energy down to small
scales where viscosity converts it to heat.  To the extent that a hot
Jupiter exchanges energy with its environment purely radiatively, so
that winds and tides are negligible in the energy balance, and to the
extent that its energy content is not secularly changing, all of the
heat absorbed from the star must ultimately be re-radiated.  In a
sense, therefore, the Carnot efficiency is always zero, but only in a
sense that would also apply to any heat engine under steady conditions
if the heat dissipated in the load were counted.

Equation \eqref{eq:entropy} allows mass elements to gain or lose heat
by radiation, and as eq.~\eqref{eq:dmech} shows, these changes can
give rise to a net gain in mechanical energy via $p\ud V$ work. But
there are no viscous terms in eqs.~\eqref{eq:euler} and
\eqref{eq:entropy} to convert mechanical energy back into heat.
Therefore, eqs.~\eqref{eq:euler}-\eqref{eq:entropy} describe a
frictionless heat engine.  The only way that such an engine could
reach steady state would by adjusting itself to zero Carnot
efficiency: that is, by arranging for the net $p\ud V$ work to vanish.
An example would be convection so efficient as to erase the vertical
temperature gradient that drives it.  Because the relevant temperature
gradient for circulation is longitudinal rather than vertical, it is
possible that the Carnot efficiency could be annulled by a $90^\circ$
phase shift between the temperature contrast and the radiative heat
exchange rather than by complete eradication of longitudinal
temperature gradients.  Something like this seems to be going on at
the deeper levels of the simulations of \cite{Showman_Fortney_etal08}:
see their Figure~3.  At the shallower levels, however, the
temperatures averaged over the day and night sides are clearly above
and below average, respectively, as might be expected from the shorter
thermal times there, so a dissipative process is required to explain
how steady state is reached at these levels.

Although the dissipation that permits a steady state is not
acknowledged in the governing equations
\eqref{eq:euler}-\eqref{eq:entropy}, it may be determined implicitly.
If the equations are appropriately averaged over height/pressure, they
reduce to shallow-water equations [more precisely,
equivalent-barotropic equations, \citet{Salby89}] with a
height-integrated ``pressure''
$\Pi\propto\Sigma^{(2\gamma-1)/\gamma}$, where $\Sigma$ is the
height-integrated density (mass per unit area).  It is well known that
the shallow-water equations permit weak solutions, i.e.
nondifferentiable ones, that contain dissipative discontinuities
(``shocks''); The amount of dissipation in the flow as it crosses one
of these discontinuities is determined by conservation laws even when
the dissipative region is unresolved \citep[see, e.g.,][]{Leveque}.
Thus any algorithm that integrates the equations stably, and that
would converge to the correct weak solution (for given initial
conditions and source terms) in the limit of infinite resolution, will
dissipate if the weak solution does, even if the algorithm doesn't use
an explicit dissipative term.  Indeed, \cite{Showman_Fortney_etal08}
call attention to ``hydraulic jumps'' in their solutions.

Unfortunately, the dissipation at hydraulic jumps may be
mathematically well-defined without being physically correct.  That
is, the dissipation isn't the same as would be obtained from ideal gas
dynamics for the same initial conditions and source function, because
eqs.~\eqref{eq:euler}-\eqref{eq:entropy} are only an approximation to
the ideal-gas equations.  In particular, the shallow-water jump
conditions conserve mass and momentum but not energy.  The energy per
unit mass plays a role in the shallow-water approximation that is
similar to that played by $(-s)$ in ideal gas dynamics: it is
conserved in smooth flow and decreases across shocks 
\citep{Landau_Lifshitz_vol6,Leveque}.
Inasmuch as eqs.~\eqref{eq:euler}-\eqref{eq:entropy} reduce to
shallow-water equations when the horizontal velocities are constant
with height, they probably destroy energy at hydraulic jumps.
Certainly there is nothing in the equations to convert the dissipated
mechanical energy to heat.

\section{Other work}

\cite{Menou_Rauscher08} adopt equations equivalent to
\eqref{eq:euler}-\eqref{eq:entropy}, but they add dissipative terms.
Some of these are diffusive terms for the temperature, vorticity, and
velocity-divergence fields.  The only terms that would appear to
dissipate kinetic energy are terms that damp the relative vorticity
and velocity divergence; in our notation, these are of the forms
\begin{align*}
  \frac{\D}{\D t}(\gradh\cross\vvh)&=\ldots 
-\frac{\gradh\cross\vvh}{\tau_{\rm fric}}\\
  \frac{\D}{\D t}(\gradh\bdot\vvh)&=\ldots 
-\frac{\gradh\bdot\vvh}{\tau_{\rm fric}}\,.
\end{align*}
For the purpose of testing their codes against previous calculations
made for the Earth's atmosphere, they take the friction time
$\tau_{\rm fric}=2\pi\Omega^{-1}$, but when they apply the model to
gaseous exoplanets, they set $\tau_{\rm fric}=\infty$, perhaps because
it is understood that these terms are intended to represent turbulent
drag against the Earth's solid surface.  Nevertheless, their
calculations saturate with subsonic wind speeds, whereas the peak
speeds in \cite{Showman_Fortney_etal08}, and also in most of the
previous calculations by Showman and his collaborators, are
supersonic.

\cite{Showman_Guillot02} integrate the primitive equations in three
dimensions.  They do not include explicit dissipative terms, but they
do remark upon a downward energy flux at the bottom of their grid,
towards the planetary interior.  It is not clear whether this is
enough of an energy sink by itself to explain the quasi-steady
circulation in the atmosphere.

\cite{Cho_etal03} and \cite{Cho_etal08} solve two-dimensional
shallow-water equations---more precisely, equivalent barotropic
equations.  There is no heat engine in these models: thermal
``forcing'' is represented by a prescribed variation in the effective
depth of the circulating layer, but in such a way that no net
work is done on the fluid in the time average.  For example, the governing
equations of \cite{Cho_etal03} for the horizontal velocity and
effective layer thickness, $h$, are (with some minor changes in notation)
\begin{align*}
  \left(\frac{\partial}{\partial t}+\vvh\bdot\gradh\right)\vvh &= 
-g\gradh(h-h_a)
+f\vk\cross\vvh,& &\mbox{(CMHS.1)}\\
  \left(\frac{\partial}{\partial t}+\vvh\bdot\gradh\right)h &= 
-\mathcal{K}\,h
\gradh\bdot\vvh\,-\frac{h-h_d}{\tau_d}\,.&&\mbox{(CMHS.2)}
\end{align*}
Here $h_a$ and $h_d$ are prescribed functions intended to represent
the forcing, and $\tau_d$ is a constant.  Integrating the righthand
side of eq.~(CMHS.1) around a closed streamline yields zero because
the accelerations parallel to the flow are gradients of scalars.
Thus, a steady-solution of these equations without dissipative terms
could be compatible with energy conservation.\footnote{This argument
  does not apply to \eqref{eq:euler} because streamlines may cross
  isobaric surfaces.  Thus if $ABB'A'$ is a hypothetical
  ``rectangular'' streamline such that the legs $AB$ and $B'A'$ lie
  entirely on neighboring isobars $p$ and $p'$, then the line integral
  of the three-dimensional gradient $\grad\Phi$ does vanish, but the
  contributions of the vertical legs $BB'$ and $A'A$ are offset by the
  pressure acceleration due to hydrostatic equilibrium
  [eq.~\eqref{eq:hydrostatic}], and these contributions do not cancel
  unless $\rho_{BB'}=\rho_{A'A}$.}  In fact, a steady solution with
$\vvh=0$ and $h=h_a$ clearly exists if $h_d=h_a$.  On the other hand,
the equations do not conserve the mass in the layer unless
$\tau_d\to\infty$, in which case (CMHS.2) can be recast as
$\partial\Sigma/\partial t+\gradh\bdot(\Sigma\vvh)=0$ for areal mass
density $\Sigma\propto h^{1/\mathcal{K}}$.

\cite{Langton_Laughlin07} integrate the shallow-water equations using
a spectral method.  They do not display any explicit dissipative term
in the momentum equations, but they mention the use of a
hyperviscosity.  \cite{Langton_Laughlin08} adopt somewhat different,
but also two-dimensional, equations with a radiative forcing that
appears to yield a nonconservative force in the momentum equation [the
term $RT\gradh\ln\rho$ in their eq.~(9)], so they may have a heat
engine.  Their only explicit mechanical dissipation is again a
hyperviscosity, $-B\nabla^4\vvh$, but the mechanical energy lost via
this term is apparently not added as heat to the evolutionary equation
for the temperature.

Not all of the work in this subject has been carried out with the
primitive, shallow-water, or equivalent-barotropic approximations.
\cite{Burkert_etal05} and \cite{Dobbs-Dixon_Lin08} use the full
ideal-gas equations, including radiative forcing terms but without the
approximation of vertical hydrostatic equilibrium.  The former work is
limited to two dimensions (radius and longitude), while the latter is
fully three dimensional.  It is not clear from their abbreviated
descriptions whether their schemes conserve total energy, but if they
do, then these schemes should be correctly converting the mechanical energy of
the circulation into heat. Even so, by neglecting explicit dissipation,
they may overestimate the flow speed.  \cite{Dobbs-Dixon_Lin08} remark upon flow
speeds up to Mach 2.7 and dissipation in shocks.

\section{Discussion}\label{sec:discussion}

Analogies between atmospheric circulation and heat engines have been
made before.  In the atmospheres of planets such as the Earth whose
rotational periods are short compared both to their orbital periods
and to the thermal times in their atmospheres\footnote{The temperature
  of the terrestrial atmosphere varies rather little over the diurnal
  cycle except at low altitude where the air easily exchanges heat
  with the ground.}, the primary temperature gradient is latitudinal,
and the atmospheric heat engine operates between the equator and the
poles, via the Hadley, mid-latitude, and polar cells.
As discussed by \citet[and references therein]{Lorenz_etal01},
the latitudinal circulation on Earth, and perhaps also on Mars and
Titan, adjusts itself approximately so as to maximize the rate of
entropy production, defined as $\dot S\equiv \dot Q_{\rm adv}(T_{\rm
  pole}^{-1}-T_{\rm equator}^{-1})$, where $\dot Q_{\rm adv}$ is the rate
of poleward heat transport by advection.  Here
$\dot S$ does not represent a secular increase in the entropy of the
atmosphere itself, since the heat and entropy brought to the poles is
ultimately radiated to space, but rather a (small) increase in the
entropy of the universe.  Maximizing $\dot S$ is not equivalent
to maximizing Carnot efficiency ($\eta_{\rm C}$), but neither is it
compatible with $\eta_{\rm C}=0$. The entropy production defined this
way would vanish for $\dot Q_{\rm adv}=0$, which would lead to the
largest difference between equatorial and polar temperatures and hence
to the largest possible $\eta_{\rm C}$; but it would also vanish if
the heat transport were so large that $T_{\rm pole}=T_{\rm equator}$,
when $\eta_{\rm C}\to 0$.  In the actual state of approximately
maximal $\dot S$, the mechanical energy of the circulation is
continuously produced by the heat engine but dissipated by turbulent
friction with the solid surface.  Since the transported heat is ultimately
radiated, a quantity related to $\dot S$ 
that could be measured in those simulations that
include radiative transport might be
\begin{equation*}
  \iint\limits_{\rm photosph.} T_{\rm ph}^{-1}\boldsymbol{F}_{\rm rad}
\bdot\ud^2\boldsymbol{A}
\end{equation*}
where $\boldsymbol{F}_{\rm rad}$ is the difference between the outward
planetary flux and the inward stellar flux, and $T_{\rm ph}$ is the
gas temperature at the photosphere.  Since the optical and infrared
photospheres do not coincide, however, a better definition might be
\begin{equation}
  \label{eq:Sdot}
  \dot S_{\rm LL} = \iiint\limits_{\rm atmos.}
\grad\bdot\left(\frac{\boldsymbol{F}_{\rm rad}}{T}\right)
\ud^3\boldsymbol{r}\,,
\end{equation}
where the subscript indicates that this is entropy production in the sense of
\citet{Lorenz_etal01}.  Normalized perhaps by the total insolation and by a
mean photospheric temperature, the quantity \eqref{eq:Sdot} might be used to
assess the degree to which the redistribution of heat is maximal.  It is not
obvious \emph{a priori} why $\dot S_{\rm LL}$ should be maximal, however.

There are at least two distinct though related ways in which
neglect of explicit dissipation could lead to erroneous results:
\begin{enumerate}
\item If mechanical energy is dissipated by numerical effects that do
  not convert it to heat, then energy is lost.  The seriousness of
  this error for predictions of infrared light curves, \emph{etc.},
  depends upon the fraction of the advected energy that is carried in
  kinetic form rather than as enthalpy.  The latter could be measured
  by integrating the two contributions to the energy flux over the
  meridional plane at the dawn and dusk terminators, for example.

\item Even when total energy is conserved or its loss is unimportant,
wind speeds and rates of redistribution of heat may be overestimated
when physical mechanisms of drag on the flow are left out.
\end{enumerate}

Terrestrial circulation is highly subsonic, so modelers presume that
errors of the first type are not serious.  This is not so obvious for
the models of hot Jupiters.  Since the simulators do not report the
balance between the kinetic and enthalpic energy fluxes, consider the
hypothesis that the wind speed $U$ at each (sufficiently deep)
pressure level adjusts itself so that the advection time $\tau_{\rm
  adv}\equiv R/U$ is proportional to the thermal time at that depth,
\begin{equation*}
  \tau_{\rm th}\equiv \frac{p}{\sigma T^4}\frac{\kappa\rho H_p^2}{c}=
\frac{\kappa  P^2 H_p}{g\sigma T^4}\,,
\end{equation*}
where $H_p\approx k_{\rm B}T/2m_pg$ is the pressure scale height.  In
an isothermal atmosphere, this hypothesis predicts
$U\propto\kappa^{-1}p^{-2}$.  The horizontal enthalpy flux would then
scale with depth as $(\kappa p)^{-1}$, and the kinetic flux as
$\kappa^{-3}p^{-5}$. Since the opacity generally increases with
pressure, the depth-integrated heat transport would be dominated by
low pressures near the photosphere, where the kinetic component is
important inasmuch as the speeds are transonic according to the
simulations.

On the other hand, if the simulations have overestimated wind speeds,
then the kinetic contribution to the energy flux may turn out to be
minor after all.  This brings us to point 2 above: What physical
mechanisms of dissipation and drag may operate in real exoplanetary
atmospheres that have not been included in the simulations?  Unlike
terrestrial planets, hot Jupiters presumably have no solid surface to
provide boundary-layer drag; a dissipative mechanism that is
distributed through the atmosphere is required.  Two important
possibilities are shocks and shear-driven turbulence (perhaps also
waves).  Shocks would occur in steady flow only if it is supersonic.
They might do little to mix the atmosphere and to resist chemical
fractionation.  If shocks dominate, this would be good for
simulations, which ought to be able to capture them, as in the
simulations of \citet{Burkert_etal05} and \citet{Dobbs-Dixon_Lin08}.
If turbulence dominates, however, then the simulations may require a
subgrid model, because a necessary characteristic of the turbulence is
that it extend to unresolvably small scales where viscosity is
effective.  The subgrid model should be based on an understanding of
the instabilities that promote the turbulence and how they saturate,
unless an empirical model can be calibrated against the well-resolved
planetary atmospheres of the solar system.

The principal obstacle to the development of three-dimensional
turbulence and a cascade to small scales is the stable stratification.
The kinetic energy available in the vertical wind shear must be
greater than the potential energy required to mix the atmosphere.  The
ratio of these energies is quantified by the Richardson number
$Ri\equiv N^2/(\partial U/\partial r)^2$, where $N$ is the
\BV frequency and $U$ is the horizontal velocity.  A
requirement for KH instability in vertical planes is that $Ri<1/4$
somewhere in the flow, at least under adiabatic and inviscid
conditions \citep{Drazin_Reid81}.  It would be helpful if numerical
researchers would report the values of $Ri$ achieved in their
simulations.  For an isothermal atmosphere with density and pressure
scale height $H_p=\cs^2/\gamma g$, the condition $Ri<\frac{1}{4}$ is
equivalent to
\begin{equation}
  \label{eq:Richardson}
  \left|\frac{\partial U}{\partial r}\right|
> \frac{2\sqrt{\gamma-1}}{\gamma}\frac{\cs}{H_p}\,,
\end{equation}
so with $\gamma\approx 7/5$ as for nondegenerate molecular hydrogen,
the flow must be transonic or else change on a scale smaller than the
scale height in order that $Ri<1/4$.  But it is possible that
instability may occur at larger Richardson number in the presence of
radiative diffusion.  Other instabilities that may be relevant are the
baroclinic instability \citep{Pedlosky_book,Vallis_book}, the
Goldreich-Schubert-Fricke instability
\citep{GoldreichSchubert67,Fricke68}, and perhaps even the
magnetorotational instability \citep{balbus91} if the shear extends to
such a depth that the atmosphere becomes substantially conducting, and
if the shear has the right sign.  It is beyond the scope of this note to
assess the importance of these instabilities for controlling the speed
of the circulation.  Nevertheless, we will discuss the GSF instability
briefly because it is illustrative and must surely occur at some level.

The GSF instability is enabled by thermal diffusion in baroclinic
atmospheres of negligible viscosity that do not rotate on cylinders,
$\partial\Omega/\partial z\ne0$, even if the atmosphere is dynamically
stable according to H{\o}lland's Criterion \citep{Tassoul78}, i.e.
baroclinically stable.  The instability is axisymmetric, and the
maximum growth rate is expressed in terms of the cylindrical radius
$\varpi$ and angular momentum per unit mass $j=\varpi^2\Omega$ by
\begin{equation*}
  \mbox{Im}(\omega)_{\max}= 
\left[\frac{1}{2\varpi^3}\left(|\grad j^2|-\frac{\partial j^2}{\partial
\varpi}\right)\right]^{1/2}
\end{equation*}
but this is approached only at wavelengths
$\lambda\lesssim2\pi(\chi/N)^{1/2}$, where $\chi=16\sigma
T^3/3\kappa\rho^2c_v\sim (c/\kappa\rho)(p_{\rm rad}/p)$ is the thermal
diffusivity, so that radiative diffusion undercuts the restoring force
of buoyancy.  Thus $\lambda\lesssim 1\,{\rm km}\,(p/{\rm
  bar})^{-1}(\kappa/{\rm cm^2\,g^{-1}})^{-1/2} (T/10^3{\rm K})^2$.
Such a wavelength would not be resolved by global simulations.  One
might expect the GSF instabilities to saturate nonlinearly at
displacements $\xi\sim\lambda$ due to nonaxisymmetric Kelvin-Helmholtz
instabilities acting on the rising and falling ``fingers,'' which
carry oppositely signed eulerian angular-velocity perturbations.
However, \citet{GoldreichSchubert67} themselves took the point of view
that the saturation occurs with displacements comparable to the
pressure scale height, and therefore provides a turbulent viscosity
$\sim\Omega H_p^2/2\pi$.  This is something best studied by local
rather than global simulations, and then represented in the latter by
a subgrid model.  \citet{Korycansky91} simulated the nonlinear outcome
of the GSF instability in the special case where $\partial j/\partial
z=0$ and $N^2>-\varpi^{-3}\partial j^2/\partial\varpi>0$.  His results
support \citet{GoldreichSchubert67}'s view of the saturation, but his
two-dimensional simulations could not have represented the
nonaxisymmetric Kelvin-Helmholtz instabilities most likely to limit
GSF modes.  \citet{Arlt_Urpin04} simulated the case $\partial
j^2/\partial\varpi>0$ and $\partial j^2/\partial z<0$ in three
dimensions, using ZEUS3D, and concluded that mixing was efficient, but
they adopted a barotropic equation of state so that their unperturbed
state had to be out of equilibrium.  \citet{Menou_Balbus_Spruit04}
added magnetic effects to the GSF analysis, but only in the linear
regime.

While the growth rate of instabilities that rely on thermal diffusion
should decrease rapidly with increasing pressure, the rate at which
radiative transfer tends to restore the stratification also decreases,
so that the outcome for the profiles of entropy and angular velocity
is unclear.  If any of these instabilities is effective at
redistributing angular momentum, then the radiatively driven
circulation may go deeper than present simulations suggest, and the
time required for the rotation profile to reach steady state may be
very long.

\vspace{20pt} We thank Adam Burrows, Kristen Menou, Jonathan Mitchell,
Geoffrey Vallis, and the Peyton-Hall astro-ph coffee klatsch for
helpful discussions.  This work was supported in part by the National
Science foundation under grant AST-0707373.

%\bibliographystyle{apj}
%\bibliography{circ}

\end{document}